# Formulation of testing gravitational redshift based on Laser Time link between China Space Station and a ground station

Rui Xu [1](ORCiD: 0000-0002-9373-9714), Wenbin Shen [1,2]*(ORCiD: 0000-0002-9267-5982), Hok Sum Fok [1](ORCiD: 0000-0003-0535-5292), Pengfei Zhang [1](ORCiD: 0000-0002-8320-293X), Lihong Li [1](ORCiD: 0000-0003-2103-1052), Lei Wang [1](ORCiD: 0000-0001-8751-4909), Kuangchao Wu [3](ORCiD: 0000-0001-6505-2870), An Ning [1](ORCiD: 0009-0004-8725-6507), Youchao Xie [1] (ORCiD: 0000-0002-2733-2166), Ziyu Shen [4](ORCiD: 0000-0002-7261-1068), Lingxuan Wang [5](ORCiD: 0000-0003-1212-023X), Yongqi Zhao [6] (ORCiD: 0000-0002-3484-950X), Kai Liu [7] (ORCiD: 0009-0008-2314-6527), Yuanjin Pan [8] (ORCiD: 0000-0002-9496-346X)

[1] Time and Frequency Observation Center (TFOC), School of Geodesy and Geomatics, Wuhan University, Wuhan 430079, China;

[2] State Key Laboratory of Information Engineering in Surveying, Mapping and Remote Sensing, Wuhan University, Wuhan 430079, China;

[3] School of Surveying and Land Information Engineering, Henan Polytechnic University, Jiaozuo, 454003, China;

[4] School of Resource, Environmental Science and Engineering, Hubei University of Science and Technology, Xianning 437100, China;

[5] School of Surveying and Mapping, Information Engineering University, Zhengzhou 450001, China;

[6] School of Geodesy and Geomatics, Wuhan University, Wuhan 430079, China;

[7] GNSS Research Center, Wuhan University, Wuhan 430079, China;

[8] School of Remote Sensing and Geomatics Engineering, Nanjing University of Information Science and Technology, Nanjing 210044, China.

***Email address: wbshen@sgg.whu.edu.cn (corresponding author)**

**Abstract:** This paper presents a high-precision gravitational redshift test using the China Space Station (CSS) Laser Time Transfer (CLT) system. We develop a comprehensive observation equation based on a $c^{-3}$ order relativistic model for space-ground clock comparison. While the CSS optical clock system is currently in the orbital debugging phase, our simulation using actual CSS orbit data achieves a gravitational redshift verification precision of $(1.8 \pm 47) \times 10^{-7}$—approximately one order of magnitude improvement better than previous experiments. Our work represents the first application of laser-based time transfer for gravitational redshift verification at such precision, and the first use of the CSS CLT link for testing this fundamental aspect of General Relativity. Unlike microwave-based methods, our laser approach avoids ionospheric effects and first-order Doppler shifts. Residual analysis identifies tropospheric delay variations and atmospheric turbulence as the primary remaining uncertainty contributors. The achieved precision enables gravitational potential difference measurements with 0.1 $m^2/s^2$ precision—offering new capabilities for both fundamental physics investigations and geodetic applications including intercontinental height transfer. This work establishes a new benchmark for high-precision tests of relativistic physics and demonstrates the transformative potential of space-based optical time transfer.

## 1 Introduction

According to Einstein's General Relativity Theory (GRT) [1], clocks at different gravitational potentials run at different rates - a prediction whose rigorous testing underpins modern physics. Since the first high-precision atomic clocks appeared, numerous experiments to test the Gravitational Red Shift (GRS) have been conducted.

In 1971, Hafele and Keating flew cesium atomic clocks eastward and westward around the globe, confirming both relativity time dilation and the GRS effects[2]. Five years later, NASA and the Massachusetts Institute of Technology (MIT) collaborated on Gravity Probe A, which lofted a hydrogen atomic clock to 10,000 km, achieving a GRS verification precision of $7\times 10^{-5}$ [3,4]. Recently, two independent data from hydrogen atomic clocks aboard Galileo navigation satellites (GSAT-0201 and GSAT-0202) were shown to reach precisions of $2.5\times 10^{-5}$ [5] and $3.1\times 10^{-5}$ [6], respectively. Using two transportable optical lattice clocks with a height difference of 450 meters in Tokyo Skytree, the precision was shown to attain $9.1\times 10^{-5}$ [7]. Despite these significant achievements, the experiment did not involve detailed modeling or systematic elimination of link and atmospheric error effects [8].

Laser Time Transfer (LTT) technology offers exceptional precision, low latency, and strong anti-interference capability, making it an ideal means for long-distance, high-precision clock synchronization [9]. Unlike traditional microwave-based time transfer technology, LTT is less affected by the atmospheric fluctuations and offers more stable signal propagation, greatly enhancing synchronization accuracy and reliability [10]. Therefore, LTT holds broad potential in metrology, fundamental physics, and space science[11].

One of the earliest LTT experiments, the Laser Synchronization from Stationary Orbit (LASSO) experiment, achieved intercontinental time synchronization with precision better than 100ps [9]. In 2008, the Time Transfer by Laser Link (T2L2) experiment onboard Jason-2 satellite further improved time transfer precision to under 100ps and reached 1ps/1000s stability [10]. The European Laser Timing (ELT) mission, proposed under the ACES framework in 2009, aims to reduce uncertainty even lower to ~25ps [11].

Following China's first successful LTT experiment on BeiDou satellites in 2008 and subsequent experiments in 2010 and 2011, Chinese researchers achieved space-ground time synchronization with 300ps precision and relative frequency stability of $10^{-14}$ [12]. More recently, the 2023 CLT experiment aboard China Space Station (CSS) has demonstrated link time stability of ~0.1ps/300s, confirming LTT's capability for high-precision space-ground clock comparison [13].

Currently, the CSS hosts a Sr optical lattice clock with a $2\times 10^{-15}/\sqrt{\tau}$ (τ in seconds) stability and is fitted with both microwave and laser time-transfer links[14–16]. A space-ground frequency comparison simulation experiment has been conducted using the CSS microwave link, demonstrating the GRS effect at $5\times 10^{-7}$ precision – an order of magnitude beyond previous experiments [17]. Meanwhile, CSS's Laser Time Transfer (CLT) system has demonstrated laboratory stability of 0.08ps/300s and 0.8ps/86400s [13], offering unprecedented accuracy low latency, and robustness against atmospheric disturbances.

In summary, the laser payload onboard the CSS provides a crucial experimental platform for gravitational redshift verification. Building upon the CSS's Sr optical lattice clock with stability of $2 \times 10^{-15} / \sqrt{\tau}$, we have developed a LTT gravitational redshift observation model accurate to the order of $c^{-3}$, based on the time transfer formalism by [18]. Building upon this foundation, the present study develops a comprehensive CLT-based observation model accurate to $c^{-3}$ order. Unlike the previous microwave-based approach [17], our laser-based method avoids ionospheric effects and first-order Doppler shifts, enabling potentially higher precision gravitational redshift verification. The innovation of this simulation experiment lies in the integration of the CSS's CLT system (with laboratory stability of 0.08ps/300s and 0.8ps/86400s) with high-precision optical clocks, pioneering the application of $c^{-3}$ order modeling for time-frequency transfer in the space environment. This approach effectively mitigates atmospheric fluctuation impacts and significantly reduces link errors through precise modeling. This technological pathway establishes a foundation for achieving higher-precision gravitational redshift verification using the CSS's laser time transfer technology, promising further validation of Einstein's General Relativity predictions.

The paper is organized as follows: Section 2 reviews the theoretical foundation of gravitational redshift and derives the $c^{-3}$ laser time-transfer observation equations; Section 3 yields an error model; Section 4 presents simulation experiments and results; Section 5 summarizes our conclusions.

## 2 Methodology

The geocentric coordinate time (TCG) serves as the fundamental time coordinate of the Geocentric Celestial Reference System (GCRS). Unlike proper time(PT), denoted by $\tau$, which is measured by physical clocks and depends on their gravitational potential and velocity, coordinate time provides a uniform time parameter for describing events in the geocentric reference frame. TCG represents the time that would be measured by an ideal clock located at infinity with zero velocity relative to the geocentric frame. Its practical realization is achieved through formulae (such as Equations 9-10) that relate the PT of atomic clocks to TCG based on their spatial coordinates and velocities in the GCRS. This distinction is crucial for our analysis: while satellite orbits and laser propagation are computed in TCG, all clock observations are recorded in PT, thus TCG must be converted accordingly[18–20,5].

As illustrated in Figure 1, ground station A transmits a laser pulse to space station S, recording the emission time $\tau_e^A$ with its local clock. Upon arrival, the onboard clock at space station S logs the reception time $\tau_b^S$, which is later relayed to the ground via the onboard communication link. Simultaneously, the laser pulse is reflected back to the ground station A by a laser retroreflector array (LRA), and the return time $\tau_r^A$ is recorded. Each laser time transfer epoch yields three key timestamp observations: $\tau_e^A$ (emission), $\tau_b^S$ (onboard reception), and $\tau_r^A$ (return) [9,21,22]. By analyzing these measurements, the gravitational potential difference observation, $\Delta U$, between the ground clock A and space-based clock S can be determined and compared against the modeled-derived gravitational potential from EGM2008. This enables verifying test of Einstein's gravitational redshift.

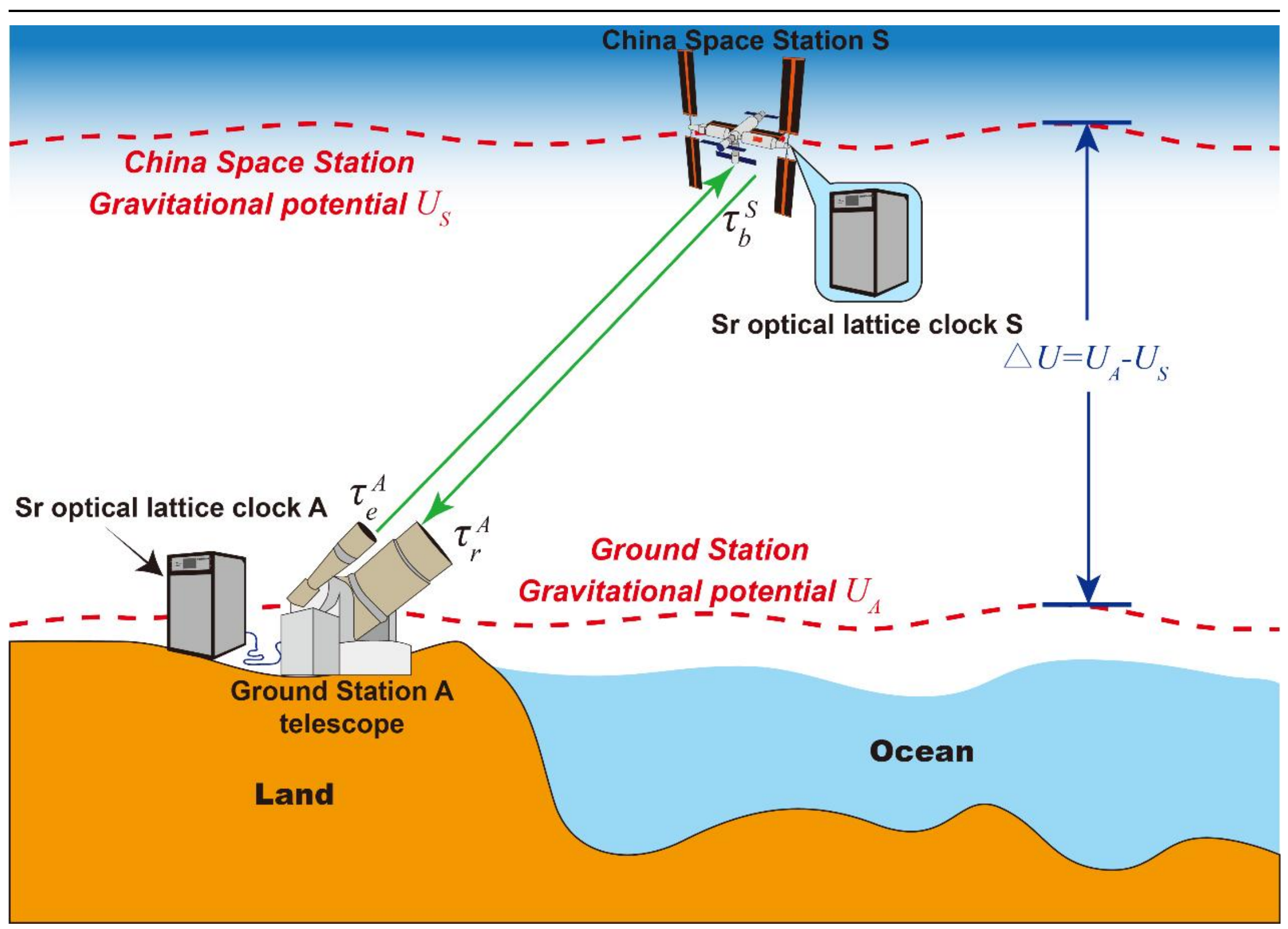

Figure 1. Testing gravitational red shift using LTT

### *2.1 LTT*

Within the GRT framework, the LTT model can be expressed as:

$$\Delta\tau^{AS} = \frac{\tau_e^A + \tau_r^A}{2} - \tau_b^S + \epsilon \tag{1}$$

where $\tau_e^A$ and $\tau_r^A$ are the signal emission and reception times recorded by the ground station clock A, $\tau_b^S$ is the time of laser arrival at the satellite recorded by the onboard clock S, and $\epsilon$ is the time delay error term [22]. Notably, the first term on the right side of the equation nearly represents the time recorded by the ground clock when the laser reaches the CSS, which in turn, the clock difference $\Delta\tau^{AS}$ can be calculated by subtracting the onboard clock observation from the ground clock observation. A detailed analysis of the error term $\epsilon$ will be discussed in Section 2.3.

## *2.2 Test of GRS*

Gravitational redshift test using clocks typically rely on measuring the relative frequency shift $\Delta\nu\ /\ \nu$ between two clocks at different gravitational potentials, where $\nu$ is the proper frequency of the clock, and $\Delta\nu$ is the frequency difference. According to general relativity, this fractional frequency difference relates to $\Delta U$ by [7,14,23]:

$$\frac{\Delta\nu_{SA}}{\nu_S} = \frac{\Delta U_{SA}}{c^2} \tag{2}$$

where the gravitational potential $U$ is defined as positive by convention in geodesy; $\Delta\nu_{SA} = \nu_A - \nu_S$; $\Delta U_{SA} = U_A - U_S$, where $\nu_A$ and $\nu_S$ are the frequencies of the ground clock and satellite clock respectively, and $U_A$ and $U_S$ are the gravitational potentials at the ground station and satellite respectively. To account for potential deviation from general relativity, a parameter $\alpha$ is introduced [5,14,24]:

$$\frac{\Delta\nu_{SA}}{\nu_S} = (1+\alpha)\frac{\Delta U_{SA}}{c^2} \tag{3}$$

where GRT holds when $\alpha = 0$. Because the gravitational potential difference between the CSS and the ground station varies with time, gravitational potential difference observations must be made epoch by epoch. Let $\Delta U(t)$ be the observed value, and $\Delta V(t)$ the corresponding true value at epoch $t$ in coordinate time. Then, equation (3) can be further rewritten as [17]:

$$\alpha = \frac{\Delta U_{SA}(t) - \Delta V_{SA}(t)}{\Delta V_{SA}(t)} \tag{4}$$

This study derives $\Delta U(t)$ from clock difference observations via LTT, using the following relationship [25]:

$$\Delta\tau_{AS}(t) = \Delta\tau_{AS}(t_0) - \int_{t_0}^{t} \frac{\Delta U_{AS}(t) + \Delta U_{TAS}(t)}{c^2} dt - \int_{t_0}^{t} \frac{\boldsymbol{v}_S^{\,2}(t) - \boldsymbol{v}_A^{\,2}(t)}{2c^2} dt$$

(5)

where $\Delta\tau_{AS}(t) = \tau_S(t) - \tau_A(t)$, and $\Delta\tau_{AS}(t_0)$ is the initial clock difference. The $\Delta U_{AS}(t)/c^2$ and $\Delta U_{TAS}(t)/c^2$ in second term on the right side accounts for the GRS effect and the Solid Earth Tide (SET) effects respectively; the third term represents the Transverse Doppler effect (TDE) difference between S and A, where $\boldsymbol{\upsilon}_A$, $\boldsymbol{\upsilon}_S$ represents the velocities of A and S in the geocentric inertial coordinate system, respectively. Considering that the term in Equation (1) represents the ground station clock observation minus the space station clock observation, the subscripts A and S in Equation (5) are interchanged to adapt it for the present study, with an additional subscript E appended to $\Delta U$ to indicate the Earth's geopotential, yielding:

$$\Delta\tau_{SA}(t) = \Delta\tau_{SA}(t_0) - \int_{t_0}^{t} \frac{\Delta U_{ESA}(t) + \Delta U_{TSA}(t)}{c^2} dt - \int_{t_0}^{t} \frac{\boldsymbol{\upsilon}_A^{\,2}(t) - \boldsymbol{\upsilon}_S^{\,2}(t)}{2c^2} dt \tag{5A}$$

Solving for the gravitational redshift term yields:

$$\int_{t_0}^{t} \frac{\Delta U_{ESA}(t)}{c^2} dt = \Delta\tau_{SA}(t_0) - \Delta\tau_{SA}(t) - \int_{t_0}^{t} \frac{\Delta U_{TSA}(t)}{c^2} dt - \int_{t_0}^{t} \frac{\boldsymbol{\upsilon}_A^{\,2}(t) - \boldsymbol{\upsilon}_S^{\,2}(t)}{2c^2} dt \tag{6}$$

To isolate the instantaneous gravitational potential difference observations epoch by epoch, we differentiate equation (6) to derive the expression for $\Delta U(t)$ in equation (4):

$$\Delta U_{ESA}(t) = -\frac{d\Delta\tau_{SA}(t)}{dt} \cdot c^2 - \Delta U_{TSA}(t) - \frac{\boldsymbol{\upsilon}_A^{\,2}(t) - \boldsymbol{\upsilon}_S^{\,2}(t)}{2} \tag{7}$$

Based on equation (7), accurate gravitational redshift estimation requires modeling the clock difference $\Delta\tau_{SA}(t)$ with fractional frequency stability at the $10^{-17}$ level. Section 2.3 provides a detailed derivation of this observation equation for $\Delta\tau_{SA}(t)$.

### *2.3 Observation Equation Formulation*

To accurately characterize the measurement process of the space-ground clock difference, we derive the observation equation from the time transfer model in vacuum [18], incorporating atmospheric effects on laser signal propagation. The core of this observation equation lies in comparing PT readings of the ground and space clocks, converting from TCG to PT via the laser propagation process.

Each PT observation includes the true value and a noise component [26], expressed as:

$$\tau = \hat{\tau} + \Delta\tau_n \tag{8}$$

where $\hat{\tau}$ is the true proper time, and $\Delta\tau_n$ is the clock noise.

Since all clock observations are recorded in PT, whereas the calculations of link errors and satellite orbits are performed in the GCRS according to the Standards of Fundamental Astronomy (SOFA)[27], it is necessary to convert all TCG-based quantities to the PT scale. The PT, $\tau$, indicated by ground clock A and the space clock S are related to TCG in non-rotating GCRS coordinates by [18,19,5]:

$$d\tau^A = \int \left( 1 - \frac{\boldsymbol{\upsilon}_A^2}{2c^2} - \frac{U_{EA} + U_{TA}}{c^2} \right) dt \tag{9}$$

$$d\tau^S = \int \left( 1 - \frac{\boldsymbol{\upsilon}_S^2}{2c^2} - \frac{U_{ES} + U_{TS}}{c^2} \right) dt \tag{10}$$

In equations (9) and (10), the integral terms account for the conversion from TCG to proper time. The first term inside the integral, $1 - \boldsymbol{\upsilon}_A^2/2c^2$, represents the special relativistic time dilation due to the ground station's motion. The second term, $-\left(U_{EA} + U_{TA}\right)/c^2$, captures the general relativistic gravitational time dilation, where $U_{EA}$ is Earth's gravitational potential and $U_{TA}$ is the tidal potential at the ground station location. These corrections ensure that the proper time recorded by the ground clock accurately reflects the relativistic effects present in the Earth's gravitational field.

Furthermore, although the theoretical midpoint $\left(\tau_e^A + \tau_r^A\right)/2$ in equation (1) assumes symmetric laser paths, the motion of the CSS and Earth's rotation break this symmetry, necessitating careful modeling.

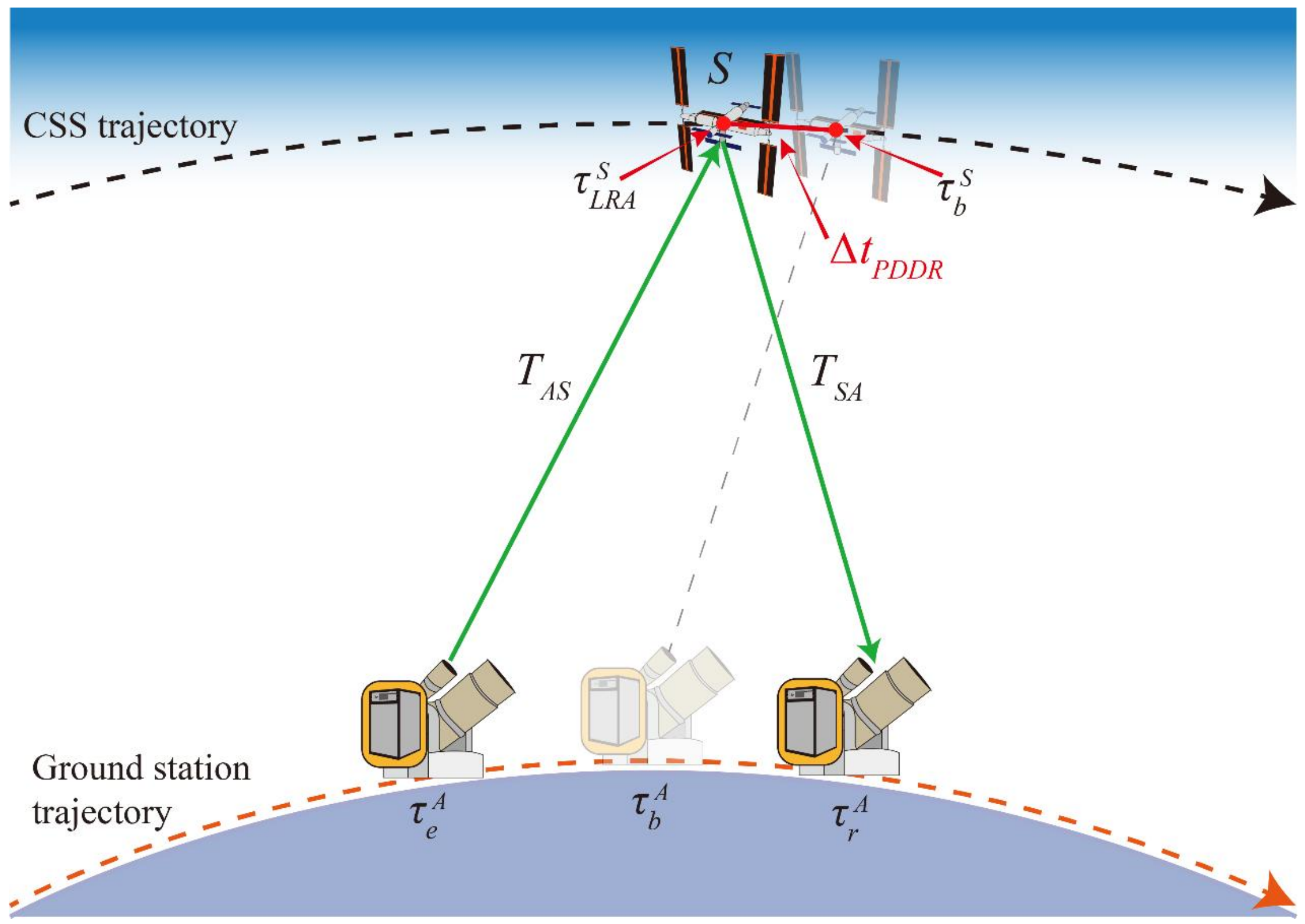

Figure 2. Principle of CLT observation sequence. In the figure, $\tau_e^A$ and $\tau_r^A$ represent the emission and return times of the laser recorded by the ground station clock A, $\tau_{LRA}^S$ represents the time when the laser reaches the satellite reflector (which cannot be recorded by clock S), $\tau_b^S$ represents the time when the laser reaches the detector recorded by the onboard clock S, $\Delta t_{PDDR}$ represents the position delay between when the laser pulse reaches the reflector and when it reaches the photon detector, and $\tau_b^A$ represents the hypothetical reading of the ground clock A when the laser reaches the detector, $T_{AS}$ and $T_{SA}$ represent the times of flight in uplink and downlink, respectively.

Figure 2 illustrates the laser transfer sequence. The pulse travels from ground station A to the CSS Laser Retroreflector Array (LRA), then to the onboard reflector (time $\tau_b^S$), with a position delay $\Delta t_{PDDR}$ that will be described in Section 3.

Assume the laser reaches the detector at ground station time $\tau_b^A$, combining these conditions with equations (8) and (9), it is written as:

$$\tau_b^A = \hat{\tau}_e^A + \Delta\tau_n^A + \int_{t_e}^{t_{LRA}+\Delta t_{PDDR}} \left(1 - \frac{\boldsymbol{\upsilon}_A^2}{2c^2} - \frac{U_{EA}+U_{TA}}{c^2}\right) dt \tag{11}$$

where $\hat{\tau}_e^A$ is the true PT emission time, $\Delta\tau_n^A$ is the noise of clock A. $t_e$ and $t_{LRA}$ are the TCG corresponding to the emission and reflector arrival time, respectively, with the explicit equation as $t_{LRA} + \Delta t_{PDDR} - t_e = T_{AS} + \Delta t_{PDDR}$, where $T_{AS}$ represents the time of flight in uplink. Then the true value $\hat{\tau}_b^A$ can be expressed as:

$$\hat{\tau}_b^A = \hat{\tau}_e^A + \int_{t_e}^{t_{LRA}+\Delta t_{PDDR}} \left(1 - \frac{\boldsymbol{\upsilon}_A^2}{2c^2} - \frac{U_{EA}+U_{TA}}{c^2}\right) dt \tag{12}$$

The onboard reading at the detector is:

$$\tau_b^S = \hat{\tau}_b^S + \Delta\tau_n^S \tag{13}$$

where $\Delta\tau_n^S$ is the noise of the onboard clock S.

Referring to equation (6), assuming no link errors, the GRS-related clock difference $\Delta\tau_{GRS}^{SA}$ is:

$$\Delta\tau_{GRS}^{SA} = \hat{\tau}_b^A - \hat{\tau}_b^S - \Delta\tau_{Dop}^{SA} - \Delta\tau_{Tide}^{SA} - \Delta\tau_0 \tag{14}$$

where $\hat{\tau}_b^A - \hat{\tau}_b^S$ is the true value of the clock difference ignoring link errors, $\Delta\tau_{Dop}^{SA}$ is the TDE, $\Delta\tau_{Tide}^{SA}$ is the SET effect, and $\Delta\tau_0$ is the initial clock difference between the two clocks related to the $\Delta\tau_{SA}(t_0)$ in equation (6).

Subsequently, the laser reflects and returns to the ground station A at time $\tau_r^A$ expressing as:

$$\tau_r^A = \hat{\tau}_e^A + \Delta\tau_n^A + \int_{t_e}^{t_{LRA}} \left(1 - \frac{\boldsymbol{\upsilon}_A^2}{2c^2} - \frac{U_{EA}+U_{TA}}{c^2}\right) dt + \int_{t_{LRA}}^{t_r} \left(1 - \frac{\boldsymbol{\upsilon}_A^2}{2c^2} - \right.$$

$$\frac{U_{EA}+U_{TA}}{c^2}\Big)dt \quad (15)$$

Substituting equations (12)-(15) into equation (1), we get:

$$\Delta\tau^{SA} = \frac{\tau_e^A+\tau_r^A}{2} - \tau_b^S = \frac{2\hat{\tau}_e^A + 2\Delta\tau_n^A + \int_{t_e}^{t_{LRA}}\left(1-\frac{\boldsymbol{v}_A^2}{2c^2}-\frac{U_{EA}+U_{TA}}{c^2}\right)dt + \int_{t_{LRA}}^{t_r}\left(1-\frac{\boldsymbol{v}_A^2}{2c^2}-\frac{U_{EA}+U_{TA}}{c^2}\right)dt}{2} - \left[\Delta\tau_n^S + \hat{\tau}_e^A + \int_{t_e}^{t_{LRA}+\Delta t_{PDDR}}\left(1-\frac{\boldsymbol{v}_A^2}{2c^2}-\frac{U_{EA}+U_{TA}}{c^2}\right)dt - \Delta\tau_{GRS}^{SA} - \Delta\tau_{Dop}^{SA} - \Delta\tau_{Tide}^{SA} - \Delta\tau_0\right] \quad (16)$$

where the conversion process between TCG and PT involves the velocity, gravitational potential, and tidal potential of ground station A. Since the TCG intervals representing the integration intervals are extremely small, these values can be considered constant. The equation above can be further simplified as follows:

$$\Delta\tau_{GRS}^{SA} = \left(\frac{\tau_e^A+\tau_r^A}{2} - \tau_b^S\right) - \left(\Delta\tau_n^A - \Delta\tau_n^S\right) - \left(1-\frac{\boldsymbol{v}_A^2}{2c^2} - \frac{U_{EA}+U_{TA}}{c^2}\right)\left(\frac{T_{SA}-T_{AS}}{2} + \Delta t_{PDDR}\right) - \Delta\tau_{Dop}^{SA} - \Delta\tau_{Tide}^{SA} - \Delta\tau_0 \quad (17)$$

where the first bracket on the right represents the original clock difference term based on observations, while the second bracket accounts for the residual clock noise. The third bracket describes the conversion relationship between TCG and the clock's PT, and the fourth bracket addresses the link error of CLT.

In the geocentric inertial coordinate system (non-rotating), assuming the TCG at the laser emission time is $t_e$, with coordinates of ground station A as $\mathbf{x}_A(t)$ and those of the CSS as $\mathbf{x}_S(t)$. The TCG transfer formula for the uplink path is defined as follows [18]:

$$T_{AS}=\frac{D_{AS}}{c}+\frac{\boldsymbol{D}_{AS}\cdot\boldsymbol{\upsilon}_S(t_e)}{c^2}+\frac{D_{AS}}{2c^3}\left(\boldsymbol{\upsilon}_S^2(t_e)+\frac{\left(\boldsymbol{D}_{AS}\cdot\boldsymbol{\upsilon}_S(t_e)\right)^2}{D_{AS}^2}+\boldsymbol{D}_{AS}\cdot\boldsymbol{a}_S\right)+$$

$$\frac{2GM_E}{c^3}ln\left(\frac{r_A+r_S+D_{AS}}{r_A+r_S-D_{AS}}\right) \tag{18}$$

where $r_A=\left|\mathbf{x}_{\mathrm{A}}(t_e)\right|$, $r_S=\left|\mathbf{x}_{\mathrm{S}}(t_e)\right|$, $\boldsymbol{D}_{AS}=\mathbf{x}_{\mathrm{S}}(t_e)-\mathbf{x}_{\mathrm{A}}(t_e)$, $D_{AS}=\left|\boldsymbol{D}_{AS}\right|$, $\mathbf{x}_{\mathrm{A}}(t_e)$ and $\mathbf{x}_{\mathrm{S}}(t_e)$ are the position vectors of ground station A and space station S at time $t_e$, respectively; $GM_E$ is the geocentric gravitational constant, $\boldsymbol{\upsilon}_S(t_e)$ is the instantaneous velocity of space station S in the geocentric inertial system at time $t_e$, and $\boldsymbol{a}_S$ is the instantaneous acceleration of space station S. To simplify the expression of equation (18), we define:

$$\begin{cases}\Delta t_{Sag-2}^{up}=\frac{\boldsymbol{D}_{AS}\cdot\boldsymbol{\upsilon}_S(t_e)}{c^2}\\ \Delta t_{Sag-3}^{up}=\frac{D_{AS}}{2c^3}\left(\boldsymbol{\upsilon}_S^2(t_e)+\frac{\left(\boldsymbol{D}_{AS}\cdot\boldsymbol{\upsilon}_S(t_e)\right)^2}{D_{AS}^2}+\boldsymbol{D}_{AS}\cdot\boldsymbol{a}_S\right)\\ \Delta t_{Sha}^{up}=\frac{2GM_E}{c^3}ln\left(\frac{r_A+r_S+D_{AS}}{r_A+r_S-D_{AS}}\right)\end{cases} \tag{19}$$

where $\Delta t_{Sag-2}^{up}$, $\Delta t_{Sag-3}^{up}$ are the $c^{-2}$ and $c^{-3}$ order Sagnac effects for the uplink path; $\Delta t_{Sha}^{up}$ is the Shapiro time delay for the uplink path. Then $T_{AS}$ can be simplified as:

$$T_{AS}=\frac{D_{AS}}{c}+\Delta t_{Sag-2}^{up}+\Delta t_{Sag-3}^{up}+\Delta t_{Sha}^{up} \tag{20}$$

Following this, the laser pulse, reflected by the CSS at time $t_b$, has its TCG transfer formula for the downlink path given as:

$$T_{SA}=\frac{D_{SA}}{c}+\frac{\boldsymbol{D}_{SA}\cdot\boldsymbol{\upsilon}_A(t_b)}{c^2}+\frac{D_{SA}}{2c^3}\left(\boldsymbol{\upsilon}_A^2(t_b)+\frac{\left(\boldsymbol{D}_{SA}\cdot\boldsymbol{\upsilon}_A(t_b)\right)^2}{D_{SA}^2}+\boldsymbol{D}_{SA}\cdot\boldsymbol{a}_A\right)+$$

$$\frac{2GM_E}{c^3}ln\left(\frac{r_A+r_S+D_{SA}}{r_A+r_S-D_{SA}}\right) \tag{21}$$

Similarly, after simplification:

$$T_{SA} = \frac{D_{SA}}{c} + \Delta t_{Sag-2}^{dn} + \Delta t_{Sag-3}^{dn} + \Delta t_{Sha}^{dn} \tag{22}$$

where $\Delta t_{Sag-2}^{dn}$, $\Delta t_{Sag-3}^{dn}$ denote the $c^{-2}$ and $c^{-3}$ order Sagnac effects for the downlink path, and $\Delta t_{Sha}^{dn}$ is the Shapiro time delay for the downlink path.

However, the above time transfer model only considers link errors in a vacuum environment. If atmospheric effects are considered, tropospheric delay errors should also be included. This paper assumes that the atmospheric refraction error for the uplink path is expressed as $\Delta t_{tro}^{up}$, and for the downlink path as $\Delta t_{tro}^{dn}$. Due to the consideration of atmospheric refraction errors, additional Sagnac effects will also be induced, set as $\Delta t_{Sag-T}^{up}$ for the uplink path and $\Delta t_{Sag-T}^{dn}$ for the downlink path. In summary, the expressions for $T_{AS}$ and $T_{SA}$ can be further improved as following.

The time transfer model outlined above only considers link errors within a vacuum environment. When atmospheric effects are incorporated, tropospheric delay errors must also be accounted for. This study assumes the atmospheric refraction error for the uplink path as $\Delta t_{tro}^{up}$, and for the downlink path as $\Delta t_{tro}^{dn}$. The inclusion of atmospheric refraction errors induces additional Sagnac effects, defined as $\Delta t_{Sag-T}^{up}$ for the uplink path and $\Delta t_{Sag-T}^{dn}$ for the downlink path. Consequently, the expressions for $T_{AS}$ and $T_{SA}$ are refined as follows:

$$T_{AS} = \frac{D_{AS}}{c} + \Delta t_{Sag-2}^{up} + \Delta t_{Sag-3}^{up} + \Delta t_{Sha}^{up} + \Delta t_{tro}^{up} + \Delta t_{Sag-T}^{up} \tag{23}$$

$$T_{SA} = \frac{D_{SA}}{c} + \Delta t_{Sag-2}^{dn} + \Delta t_{Sag-3}^{dn} + \Delta t_{Sha}^{dn} + \Delta t_{tro}^{dn} + \Delta t_{Sag-T}^{dn} \tag{24}$$

Moreover, the impact of atmospheric turbulence on the laser link is non-negligible. Assume the total atmospheric turbulence impact on the two-way laser link is $\Delta_{Tur}$ (details in Section 3). Additionally, system delays occur in various hardware components of both the ground and space stations. The overall system error in the LTT model (equation 1) is denoted as $\Delta_{sys}$ (details in Section 3). Given that ionospheric refraction effects on laser signal propagation are negligible [28], this study excludes ionospheric refraction impacts. By substituting equations (23) and (24) into equation (17) and incorporating $\Delta t_{Tur}$ and $\Delta t_{sys}$, the error elimination model for gravitational redshift clock difference observation in the two-way laser link is derived as:

$$\Delta\tau_{GRS}^{SA} = \left(\frac{\tau_e^A + \tau_r^A}{2} - \tau_b^S\right) - \left(\Delta\tau_n^A - \Delta\tau_n^S\right) - \left(1 - \frac{\boldsymbol{\upsilon}_A^{\;2}}{2c^2} - \frac{U_{EA} + U_{TA}}{c^2}\right)$$
$$\left(\begin{array}{c} \frac{D_{SA} - D_{AS}}{2c} + \frac{\Delta t_{Sag-2}^{dn} - \Delta t_{Sag-2}^{up}}{2} + \frac{\Delta t_{Sag-3}^{dn} - \Delta t_{Sag-3}^{up}}{2} \\ + \frac{\Delta t_{Sha}^{dn} - \Delta t_{Sha}^{up}}{2} + \frac{\Delta t_{tro}^{dn} - \Delta t_{tro}^{up}}{2} + \frac{\Delta t_{Sag-T}^{dn} - \Delta t_{Sag-T}^{up}}{2} \\ + \Delta t_{PDDR} + \Delta t_{Tur} + \Delta t_{sys} \end{array}\right)$$
$$-\Delta\tau_{Dop}^{SA} - \Delta\tau_{Tide}^{SA} - \Delta\tau_0 \qquad (25)$$

where $\Delta\tau_{GRS}^{SA}$, $\Delta\tau_{Dop}^{SA}$, $\Delta\tau_{Tide}^{SA}$, $\Delta\tau_0$ correspond to $\int_{t_0}^{t} \frac{\Delta U(t)}{c^2} dt$, $\int_{t_0}^{t} \frac{\Delta\boldsymbol{\upsilon}^2(t)}{2c^2} dt$, $\int_{t_0}^{t} \frac{\Delta U_T(t)}{c^2} dt$, and $\Delta\tau(t_0)$ in equation (6), respectively, with $\Delta\tau(t_0)$ as a constant term. In practice, clock noise cannot be eliminated during data processing, and only PT observations inclusive of clock noise are obtainable. Consequently, the second bracket on the right side of equation (25) is omitted. In conclusion, the final CLT clock difference $\Delta\tau$ with the augmented link error model is represented as:

$$\Delta\tau_{SA} = \left(\frac{\tau_e^A + \tau_r^A}{2} - \tau_b^S\right) - \left(1 - \frac{\boldsymbol{\upsilon}_A^{\;2}}{2c^2} - \frac{U_{EA} + U_{TA}}{c^2}\right)$$

$$\begin{pmatrix} \frac{D_{SA}-D_{AS}}{2c} + \frac{\Delta t^{dn}_{Sag-2}-\Delta t^{up}_{Sag-2}}{2} + \frac{\Delta t^{dn}_{Sag-3}-\Delta t^{up}_{Sag-3}}{2} \\ + \frac{\Delta t^{dn}_{Sha}-\Delta t^{up}_{Sha}}{2} + \frac{\Delta t^{dn}_{tro}-\Delta t^{up}_{tro}}{2} + \frac{\Delta t^{dn}_{Sag-T}-\Delta t^{up}_{Sag-T}}{2} \\ + \Delta t_{PDDR} + \Delta t_{Tur} + \Delta t_{sys} \end{pmatrix} \tag{26}$$

In observation equation (26), the differences in propagation geometric paths, $c^{-2}$ order Sagnac effect, $c^{-3}$ order Sagnac effect, Shapiro effect, tropospheric delay, and additional Sagnac effect due to tropospheric delay in the third bracket on the right side of the equation are all caused by the inconsistency between the emission and reflection paths.

By integrating equations (4), (7), and (26), we derive the CLT model for testing gravitational redshift with precision up to $c^{-3}$:

$$\alpha = \left\{ \begin{array}{c} -d \left[ \begin{array}{c} \left( \frac{\tau^A_e + \tau^A_r}{2} - \tau^S_b \right) - \left( 1 - \frac{\boldsymbol{v}_A^2}{2c^2} - \frac{U_{EA}+U_{TA}}{c^2} \right) \\ \begin{pmatrix} \frac{D_{SA}-D_{AS}}{2c} + \frac{\Delta t^{dn}_{Sag-2}-\Delta t^{up}_{Sag-2}}{2} + \frac{\Delta t^{dn}_{Sag-3}-\Delta t^{up}_{Sag-3}}{2} \\ + \frac{\Delta t^{dn}_{Sha}-\Delta t^{up}_{Sha}}{2} + \frac{\Delta t^{dn}_{tro}-\Delta t^{up}_{tro}}{2} + \frac{\Delta t^{dn}_{Sag-T}-\Delta t^{up}_{Sag-T}}{2} \\ + \Delta t_{PDDR} + \Delta t_{Tur} + \Delta t_{sys} \end{pmatrix} \end{array} \right] \cdot \frac{c^2}{dt} \\ -\Delta U_{TSA}(t) - \frac{\boldsymbol{v}_A^2(t) - \boldsymbol{v}_S^2(t)}{2} - \Delta V(t) \end{array} \right\} / \Delta V(t) \tag{27}$$

## 3 Error sources

### *3.1 Errors Affecting Laser Propagation Time*

In precise time transfer, multiple error sources affect laser propagation time and must be accurately modeled to achieve high precision. The Sagnac effect, which arises from the relative motion between receiver and transmitter, introduces significant correction terms in time transfer equations. For space stations at approximately 400 km altitude (like the ISS and CSS), this effect can introduce delays of up to 200 ns ($c^{-2}$ term) and 5 ps ($c^{-3}$ term) at low elevation angles, making it the dominant delay source in the laser link[18].

The Shapiro delay, arising from spacetime curvature, causes electromagnetic signals to propagate more slowly in gravitational fields than in flat spacetime [29]. For the ISS, this delay is approximately 2 ps at zenith and up to 11 ps at lower elevation angles. [18].

Tropospheric delay results from refraction effects in Earth's atmosphere, with contributions from both hydrostatic (dry air pressure) and wet (water vapor) components. This delay depends on the signal's zenith angle and is corrected through mapping functions [30,31]. The total slant tropospheric delay can be calculated as [32]:

$$\Delta t_{tro}^{up} = \frac{\Delta L_h^z \cdot MF_h(\varepsilon) + \Delta L_w^z \cdot MF_w(\varepsilon)}{c} \tag{28}$$

where $\Delta L_h^z$ is the zenith hydrostatic delay component, $\Delta L_w^z$ is the zenith wet delay component, $MF_h$ and $MF_w$ are the hydrostatic and wet components of the mapping function, respectively. This study employs the GPT3 (Global Pressure and Temperature 3) 1°×1° grid model and VMF3 (Vienna Mapping Functions 3) mapping function [33], with the mapping function expressed as[30,31]:

$$MF(\varepsilon) = \frac{1 + \frac{a}{1 + \frac{b}{1 + c}}}{\sin(\varepsilon) + \frac{a}{\sin(\varepsilon) + \frac{b}{\sin(\varepsilon) + c}}} \tag{29}$$

with coefficients a, b, and c provided by the VMF3 model [33]; $\varepsilon$ is the elevation angle.

Additionally, tropospheric refraction introduces an additional Sagnac effect due to satellite motion and Earth's rotation, with the correction given as: [13,34]:

$$\Delta t_{Sag-T}^{up} = \Delta t_{tro}^{up} \cdot \frac{\boldsymbol{D}_{AS} \cdot \boldsymbol{v}_S(t_e)}{D_{AS} \cdot c} \tag{30}$$

Position inconsistencies between the photon detector and reflector introduce measurable delays [13]. Assuming zero Euler angles and Earth-facing laser payload with an offset d = 126.305 mm, this delay can be expressed as [13]:

$$\Delta t_{PDDR} = \frac{d \cdot sin(\theta)}{c} \quad (31)$$

where θ is the incidence angle of the laser arriving at the space station's laser payload given by [35]:

$$sin(\theta) = \frac{R_E}{h + R_E} \cdot sin(\phi + 90°) \quad (32)$$

where $R_E$ is the average Earth's radius, $h$ is the CSS altitude, and $\phi$ is the telescope elevation angle.

Atmospheric turbulence significantly affects laser time transfer stability through beam spread, wander, and scintillation [36]. These effects are characterized by the Kolmogorov spectra [37]:

$$\Phi_n(\kappa, z) = 0.033C_n^2(z)\kappa^{-11/3} \quad (33)$$

where $\kappa$ is the spatial frequency (in rad/m) and $C_n^2(z)$ is the refractive index structure constant along the propagation path $z$.

Using Taylor's frozen flow hypothesis, spatial frequency $\kappa$ can be converted to temporal frequency $f$ [38]:

$$\kappa = \frac{2\pi f}{V} \quad (34)$$

For the CSS's two-way time transfer links with path asymmetries, the time noise fluctuations in terms of power spectrum can be expressed as [39]:

$$\sigma_{\tau,2Way}^{2} = \frac{2\pi^2}{c^2} L \int_0^{\infty} \frac{2\pi f}{V} \Phi_n\left(\frac{2\pi f}{V}\right)\left[1 - J_0\left(\frac{2\pi f}{V} \cdot d\right)\right] \mathrm{d}f \quad (35)$$

where $d$ is the path separation. For the small-angle $\kappa d$, the Bessel function $J_0(\kappa d)$ at low frequencies simplifies to [39]:

$$J_0(\kappa d) \approx 1 - \frac{(\kappa d)^2}{4} \quad (36)$$

yielding the low-frequency time noise spectrum:

$$S_x(f)_{PR} \propto f^{-2/3} \tag{37}$$

Furthermore, time instability can then be calculated by integrating the time noise spectrum $S_x(f)$ [39] as:

$$\sigma_x^2(\tau) = \frac{8}{3k^2}\int_0^\infty \left[\frac{\sin^3(\pi f \tau)}{\sin(\pi f \tau_0)}\right]^2 S_x(f)\mathrm{d}f \tag{38}$$

where $\sigma_x^2$ is the time variance, used to measure time instability or changes in time offset within the observation time interval $\tau$, serving as a measure of time stability; $\tau$ is the observation time interval, $\tau_0$ is the sampling interval, $k$ is the sampling interval in steps of $\tau_0$, satisfying $\tau = k\tau_0$, $S_x(f)$ is the time frequency noise spectrum, and $f$ is the time frequency.

System errors also significantly impact laser time transfer links, caused by delay inaccuracies in various hardware components. The overall system error can be evaluated using [40]:

$$\sigma_{sys}^2 = \frac{\sigma_e^2}{4} + \frac{\sigma_r^2}{4} + \sigma_b^2 \tag{39}$$

Similarly, according to the law of error propagation, the relationship between system errors of the above observables and individual hardware errors can be expressed as:

$$\sigma^2 = \sigma_1^2 + \sigma_2^2 + \dots \tag{40}$$

where $\sigma$ is the system errors of the above observables ($\sigma_e$, $\sigma_b$, $\sigma_r$), and $\sigma_1$, $\sigma_2$ are individual hardware errors.

According to CLT laboratory metrics, the system error for the ground station emission time ($\sigma_e$) is estimated at 7.07 ps, the onboard arrival time error ($\sigma_b$) is 39 ps, and the ground station echo reception time error ($\sigma_r$) is 33.5 ps. Combining these errors, the overall system error of laser time transfer is estimated at 42.6 ps [35].

### *3.2 Errors Affecting Clock Operation*

SET represent the elastic deformation of Earth caused by celestial tidal forces, primarily from the Moon and Sun. The influence of celestial tidal potential on Earth's external gravitational potential manifests in three key aspects: direct impact on the external gravitational potential; changes due to Earth's mass redistribution; and changes resulting from vertical displacement of ground stations. According to IERS Conventions (2010), SET is the most significant tidal effect, with surface displacement impacts far exceeding other tidal types such as ocean and polar tides [28]. SET can produce maximum displacement amplitudes of several tens of centimeters [41], while other tidal phenomena typically cause displacements of only a few centimeters [42]. Given SET's significant influence, precise geophysical and astronomical research must account for its displacement and gravitational potential disturbances to ensure accurate observations and calculations.

The direct influence of celestial tidal potential on the Earth's external gravitational potential can be expressed by changes in spherical harmonic coefficients [28]:

$$\Delta\bar{C}_{nm} - i\Delta\bar{S}_{nm} = \frac{1}{2n+1}\sum_{j=2}^{10}\frac{GM_j}{GM_\oplus}\left(\frac{R_e}{r_j}\right)^{n+1}\bar{P}_{nm}\left(sin\Phi_j\right)e^{-im\lambda_j} \tag{41}$$

where $GM_j$ represents the gravitational constant of the tide-generating body $j$; $r_j$ is the geocentric distance of the tide-generating body; $j = 2\sim 10$ represent the Moon (n=2,3), Sun, Mercury, Venus, Mars, Jupiter, Saturn, Uranus, Neptune (n=2); $\Phi_j$ is the geocentric latitude of the tide-generating body in the Earth-fixed coordinate system; $\lambda_j$ is the longitude of the tide-generating body in the Earth-fixed coordinate system (Greenwich sidereal time).

TDE and GRS are collectively referred to as relativistic effects in the field of Laser Time Transfer (LTT). Since the purpose of this research is to extract clock offset information induced by GRS from laser signals, it is necessary to consider the residual amount of TDE and its impact on optical clock observations.

Remote clock comparison, a key step in verifying gravitational redshift, fundamentally relies on high-precision atomic clocks to measure time or frequency. The performance (stability) of atomic clocks directly determines measurement precision, while measurement noise primarily originates from two sources: the conversion process from microwave signals to electrical signals, and the inherent noise in frequency signal measurements by atomic clocks.

Atomic clock noise, determined by the inherent characteristics of oscillators and measurement systems, significantly affects clock stability. The noise distribution can be represented as a linear superposition of five typical noise types [43–45]:

$$y(t) = \sum_{\alpha=-2}^{2} z_\alpha(t) = z_{-2}(t) + z_{-1}(t) + z_0(t) + z_1(t) + z_1(t) \tag{42}$$

where $\alpha$ from 2 to -2 corresponds to the following noise components: White Phase Modulation noise (WPM), Flicker Phase Modulation noise (FPM), White Frequency Modulation noise (WFM), Flicker Frequency Modulation noise (FFM), and Random Walk Frequency Modulation noise (RWFM) [45,46].

These noises can be characterized in the frequency domain through the power spectral density (PSD) of the normalized frequency difference y(t). The power-law noise model distinguishes different noise types, with its PSD represented as [47,48]:

$$S_y(f) = \sum_{\alpha=-2}^{2} h_\alpha f^\alpha \tag{43}$$

where $h_\alpha$ is a frequency-independent intensity coefficient, and $\alpha$ represents the exponent of different noise types.

The spectral density of the normalized frequency difference $y(t)$ has the following relationship with the spectral density of the phase difference $x(t)$ after integration [49]:

$$S_x(f) = S_y(f) \cdot (2\pi f)^{-2} \tag{44}$$

## 4 Simulation experiment

Currently, the CSS has been completed, but the optical clock system is still in the orbital debugging stage and cannot perform remote time-frequency comparison, so it is not possible to conduct actual experiments to verify gravitational redshift based on our research. Nevertheless, we can still simulate the content of this research using already published relevant parameters and indicators to verify the feasibility and precision of using CLT to test gravitational redshift.

### *4.1 Experimental Setup*

In this simulation experiment, we selected the Xi'an Laser Station (XLS) as the ground station, with CSS orbit and CLT link-related parameters as shown in Table 2. The simulation specifically focused on September 2023, as this period provided optimal observation geometry with the CSS passing over XLS multiple times daily at various elevation angles, allowing comprehensive assessment of atmospheric effects and link performance across different viewing conditions.

The gravitational potential at XLS (569.247 m) was precisely determined using the EGM2008 model, yielding a value of 62556688.86765$m^2/s^2$. This accurate determination is crucial for gravitational redshift calculations, as the potential difference between space and ground directly determines the magnitude of the redshift effect being tested.

**Table 2. CSS and CLT link-related parameters**

| Parameter | Value |
| --- | --- |
| orbit altitude | 340~450km |
| orbit inclination | 41~42° |
| Xian coordinate | 34°8′33.04071N, 108°59′47.66370″E, 569.247 m |
| Cutoff elevation angle | 15° |
| Laser wavelength | 532nm |

### *4.2 Simulation Data Construction*

For this simulation experiment, we developed a comprehensive approach to model the CSS-to-ground clock comparison link. The simulation process integrated orbit data, clock performance, and various error sources to create a realistic testing environment.

We used the actual CSS orbit data published on the official China Manned Space website as the true values, adding orbit error parameters (position: 10 cm; velocity: 1 mm/s) to simulate realistic observation conditions (see Table 3). The simulated orbit data were then combined with XLS coordinates to model the link geometry and associated errors.

We simulated the performance of both the space-based and ground-based optical clocks using the Python library Allantools. The stability of the CSS optical clock was set to $2\times 10^{-15}/\sqrt{\tau}$, while the ground station clock stability was set to $1\times 10^{-15}/\sqrt{\tau}$, reflecting the better operating environment on the ground. Figure 3 demonstrates that the stability of both clocks improves with increasing integration time, from the short-term ($\tau$ = 1s) level of $10^{-15}$ to the long-term ($\tau$ = 100000s) level of $10^{-17}$, with the ground clock performing slightly better due to fewer environmental disturbances.

We systematically modeled various error sources affecting the CLT. The accuracy of various error models are listed as follows (see Table 3):

(i) Gravitational Potential: Using EGM2008 with accuracies of 0.3 $m^2/s^2$ (CSS) and 0.5 $m^2/s^2$(XLS) [50]

(ii) Tropospheric Delay: Modeled with 2.64±3.15mm accuracy [33]

(iii) Solid Earth Tides: Modeled with 0.1 $m^2/s^2$accuracy [28]

(iv) Atmospheric Turbulence: In the CLT, the atmospheric turbulence-induced time deviation is <1fs over 300s [13].

(v) System Errors: System stability set to 0.1 ps@300s and overall error to 43ps [35]

Based on these simulations, we constructed the observation values $\tau_e^A$, $\tau_b^S$, and $\tau_r^A$ from equation (1) and applied equation (26) to eliminate various error terms, allowing us to isolate and evaluate each error source's impact.

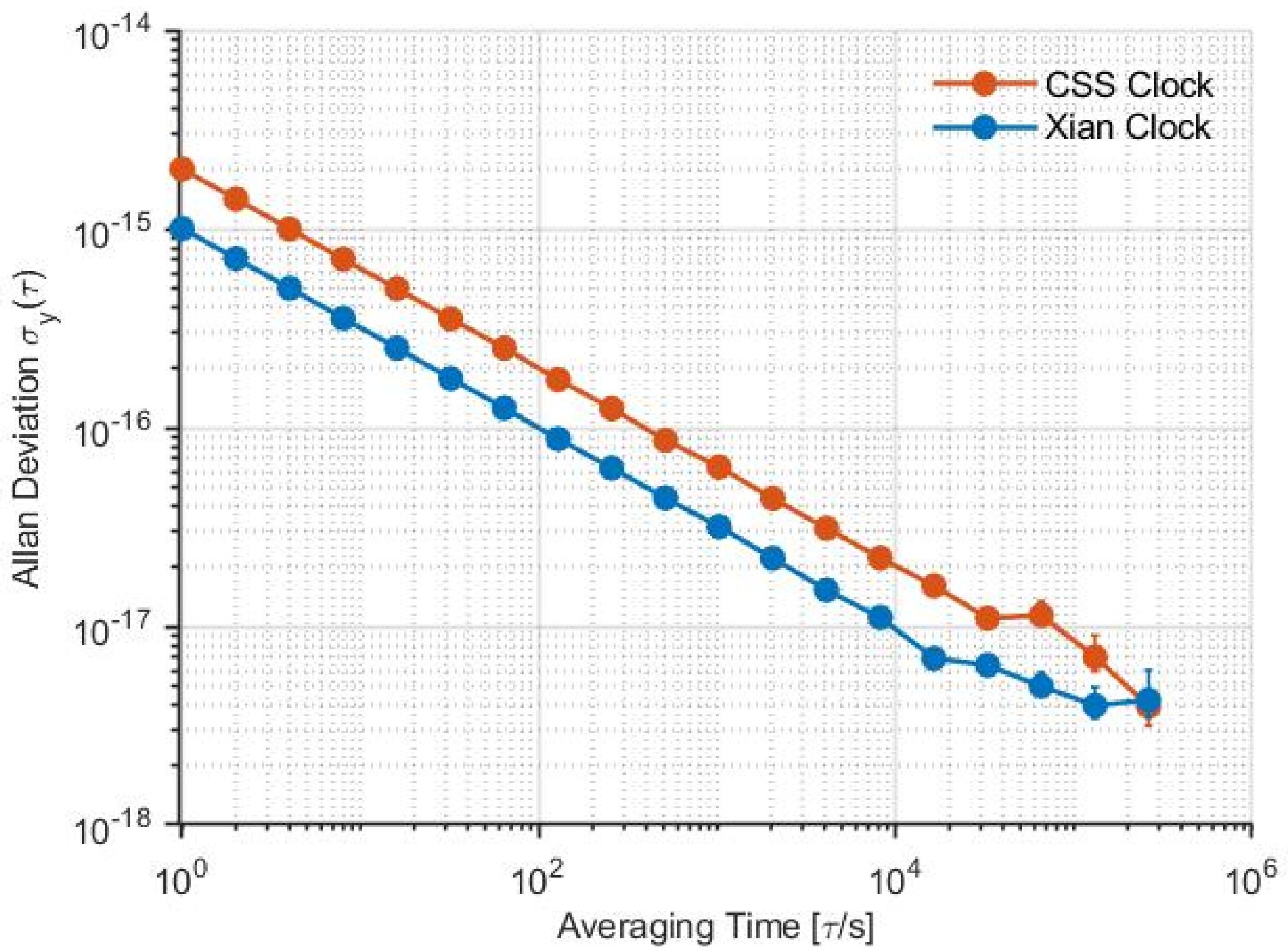

**Figure 3. Allan variance evaluation results of space-based optical clock and Xi'an station optical clock noise simulation**

**Table 3. Model accuracy for various error sources**

| **Error Model** | **Accuracy** |
|---|---|
| Ground optical clock stability | $1\times 10^{-15}/\sqrt{\tau}$ |
| Space-based optical clock stability | $2\times 10^{-15}/\sqrt{\tau}$ |
| Tropospheric delay | 2.64(±3.15) mm |
| CSS orbit error | Position: 10 cm; Velocity: 1 mm/s |
| EGM2008 | CSS： 0.3 $m^2/s^2$； XLS： 0.5 $m^2/s^2$ |
| Solid earth tide | 0.1 $m^2/s^2$ |
| Atmospheric Turbulence | <1fs @300s |
| System Errors | overall error: 43ps; stability: 0.1 ps@300s |

### *4.3 Residual Analysis*

In laser-based time synchronization and transfer systems, propagation time delays constitute a primary error source. However, as Equation (7) demonstrates, gravitational potential observations and gravitational redshift testing precision depend not on the clock offset itself but on its derivative. Our study therefore examines how various time delay sources influence the stability of the clock comparison link, requiring evaluation metrics beyond conventional LTT performance measures.

Building on the theoretical framework established in Chapter 2.3, we analyzed how various error sources affect link stability after model correction. Figure 4 presents an Allan deviation analysis of these residual impacts across different averaging times. Higher Allan variance values indicate lower stability and greater error fluctuation, while lower values reflect higher stability and more consistent performance.

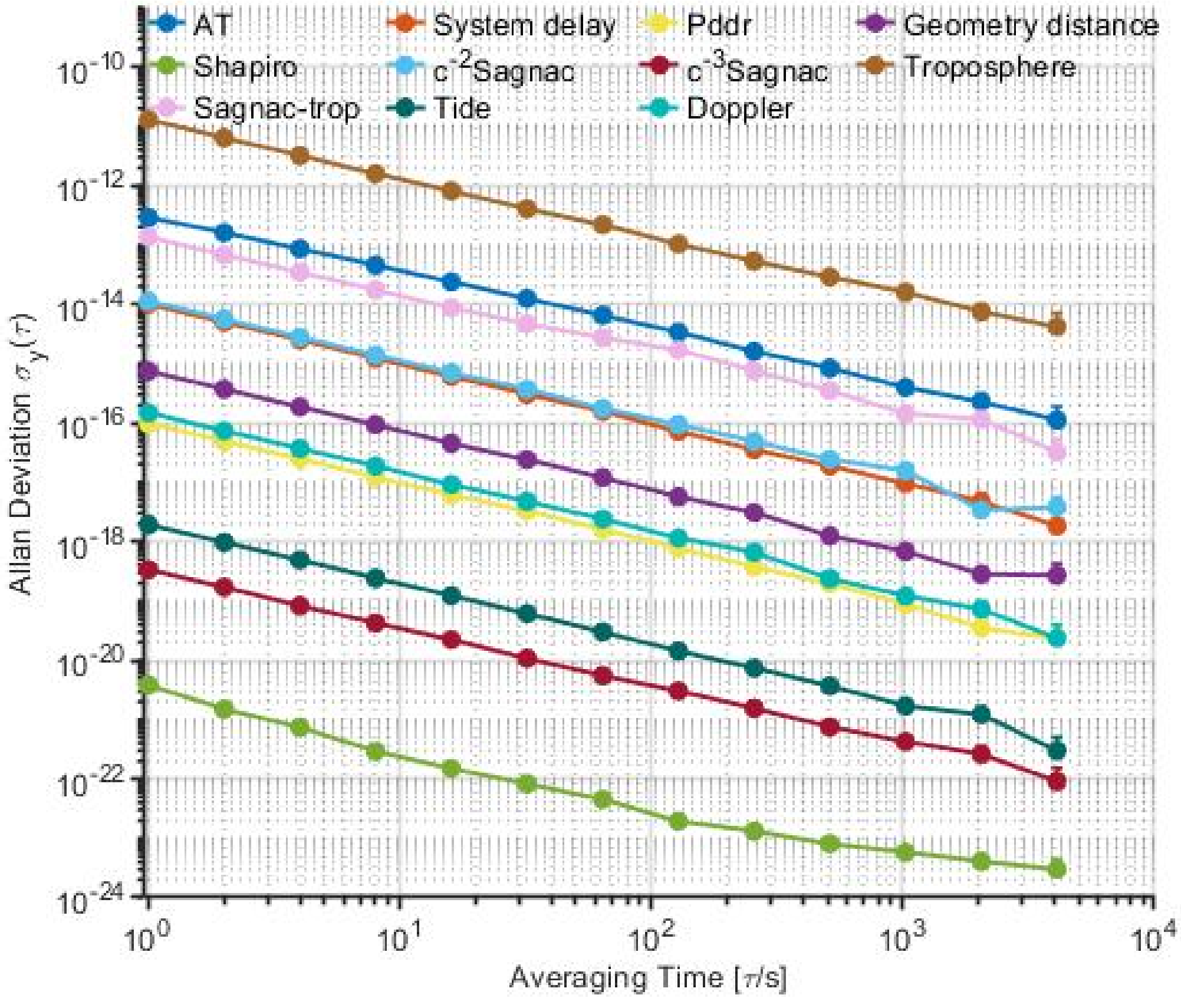

Figure 4. Impact of residuals from various error sources on clock observation stability.

The figure illustrates residual effects after applying model corrections. Note that atmospheric turbulence (AT) and system delay were directly simulated using their stability specifications, as deterministic correction models do not exist for these stochastic phenomena. All residuals demonstrate improved stability (decreasing Allan deviation) with increased averaging time, indicating diminishing impact on the clock comparison link.

Tropospheric delay emerges as the most significant residual contributor, with an Allan deviation of $4.17\times 10^{-15}$ at 4000 seconds. Despite employing sophisticated GPT3 grid modeling and VMF3 mapping functions, tropospheric inhomogeneities remain difficult to characterize fully, especially at low elevation angles. This residual dominates overall link instability but gradually diminishes over extended averaging periods as atmospheric variations even out.

Atmospheric turbulence represents the second largest impact, showing an Allan deviation of $1.12\times 10^{-16}$ at 4000 seconds. We simulated this effect directly based on established stability metrics rather than applying correction models. While causing time deviations below 1 fs over 300 seconds and exhibiting poor short-term stability, turbulence effects improve markedly with longer averaging times due to their inherently random fluctuation patterns.

The Sagnac-troposphere coupling effect exhibits a residual Allan deviation of $3.2\times 10^{-17}$ at 4000 seconds. This phenomenon stems from interactions between Earth's rotation and tropospheric delay asymmetries in uplink and downlink paths. Though initially significant, its impact diminishes rapidly with increased averaging time.

The $c^{-2}$ Sagnac effect residual reaches $3.79\times 10^{-18}$ at 4000 seconds. Our two-way measurement approach effectively cancels most of this contribution, leaving only minimal residual impact that quickly stabilizes with longer averaging periods.

Like atmospheric turbulence, we simulated system delay directly using CLT design specifications rather than applying correction models. The simulated system delay shows improvement to $1.83\times 10^{-18}$ at approximately 4000 seconds. Our simulation incorporated an overall system error of 43 ps with 0.1 ps@300 s stability, matching CLT design parameters. Given the brief CSS transit window of roughly 5 minutes, each observation segment's stability aligns with these specifications.

Geometry distance residuals, stemming from orbit determination accuracy and path inconsistencies between upward and downward links, show moderate initial impact but steadily decrease with longer averaging times, reaching approximately $10^{-19}$ at 4000 seconds.

The Position Delay of Detector and Reflector (PDDR) residual reaches $2.34\times 10^{-20}$ at 4000 seconds. This effect primarily depends on link geometry modeling accuracy and shows consistent improvement with increased averaging time.

Transverse Doppler Effect (TDE) residuals stabilize at $2\times 10^{-20}$ after 4000 seconds. Though not among the most significant error sources, this relativistic effect requires proper compensation for high-precision measurements [18,51].

Solid Earth tides (SET) residuals stabilize at $3\times 10^{-22}$ after 4000 seconds. Despite their deterministic periodic nature, these effects demand precise modeling due to their complex spatiotemporal variations.

The $c^{-3}$ Sagnac effect shows negligible impact with residuals of $9.08\times 10^{-23}$ at 4000 seconds—approximately five orders of magnitude smaller than its $c^{-2}$ counterpart.

The Shapiro delay contributes the smallest residual effect with an Allan deviation of merely $2.99\times 10^{-24}$ at 4000 seconds, making it inconsequential for both short and long-term measurements.

When comparing our CLT-based approach with [14] tri-frequency combination (TFC) method, several key differences emerge in error source handling. Sun's method focused primarily on mitigating first-order Doppler effects through microwave frequency combinations, achieving residual stabilities no better than the $10^{-14}$ range. Our laser-based approach, by contrast, delivers significantly better residual stabilities, with most effects below $10^{-17}$ after 1000 seconds of averaging.

The TFC method required three separate frequency links, whereas our CLT

approach achieves superior performance with a single wavelength. Additionally, Sun's approach suffered from inherent microwave link instability, particularly in tropospheric delay modeling, where residuals remained at the $10^{-13}$- $10^{-14}$ level even after correction [14].

Our comprehensive modeling of relativistic effects, including higher-order Sagnac terms and Shapiro delay, represents a significant advancement over previous approaches. While both our model and [14] extend to $c^{-3}$ order terms, our laser-based implementation demonstrates more effective error mitigation, as evidenced by the extremely low residual levels shown in Figure 4.

Based on our analysis, we can categorize the residual errors into three distinct tiers according to their impact on link stability at 4000 seconds:

(1) Primary contributors ( $10^{-15}$- $10^{-16}$): Tropospheric delay and atmospheric turbulence

(2) Secondary contributors ( $10^{-17}$- $10^{-18}$): Sagnac-troposphere coupling, $c^{-2}$ Sagnac effect, system delay, and geometry distance

(3) Minimal contributors ( $10^{-20}$ and below): PDDR, transverse Doppler effect, solid Earth tides, $c^{-3}$ Sagnac effect, and Shapiro delay

This hierarchical understanding guides our optimization strategy, focusing efforts on mitigating primary contributors while ensuring lower-tier effects remain sufficiently controlled for high-precision gravitational redshift testing.

### *4.4 Application and Validation of CLT*

Our CLT method offers distinct advantages over traditional microwave-based approaches for gravitational redshift testing. Compared to the work of [14] and [17] , our 532 nm laser signals provide fundamentally higher absolute frequency precision than X-band (8 GHz) and Ka-band (32 GHz) microwave signals at equivalent relative stability. Laser signals also largely bypass ionospheric effects—a persistent challenge in microwave approaches that required Sun's complex tri-frequency combinations. As [52] demonstrated, optical frequencies experience ionospheric delays roughly $10^{-7}$ times smaller than microwave frequencies.

The naturally collimated laser beams we employ significantly reduce multipath effects and interference that have consistently troubled microwave approaches [53]. Our CLT design benefits from nearly symmetrical uplink and downlink paths, enabling more effective common-mode error cancellation. This symmetry proves especially valuable for Sagnac effect compensation, yielding residuals about two orders of magnitude smaller than those reported by [14]. We've carefully incorporated all significant relativistic effects, building on the framework established by [54] and further developed by [55].

To validate our method, we implemented equation (26) to tackle the error terms outlined in Chapter 3. Our error model successfully addressed various sources of uncertainty, with hardware delay stability reaching an impressive $1.83\times 10^{-18}$ at 4000 seconds averaging time—just one aspect of our comprehensive error reduction approach. This hardware performance approaches the theoretical boundaries for optical frequency comparisons that [56] discussed and marks a clear advancement over the hardware constraints inherent in microwave systems. Overall, our CLT method substantially outperforms [14] TFC approach, which achieved a best-case stability of only $4.9\times 10^{-15}$ after corrections. These improvements largely stem from the inherent advantages of laser technology combined with our more sophisticated relativistic modeling, enabling us to measure gravitational redshift with unprecedented precision compared to traditional microwave techniques.

The effectiveness of our error mitigation varies across different sources. Our tropospheric modeling reduces initial effects from roughly $10^{-11}$ to $4.17\times 10^{-15}$ at 4000 seconds—a four-order-of-magnitude improvement that substantially outperforms microwave approaches, which typically achieve only two to three orders of magnitude improvement [57]. Our two-way measurement approach cuts the $c^{-2}$ Sagnac effect from about $1\times 10^{-14}$ to $3.79\times 10^{-18}$ at 4000 seconds, exceeding the capabilities of one-way frequency transfer methods described by [58]. We've reduced hardware-related delays to the $10^{-18}$ level, comparable to the best ground-based optical clock comparisons [59].

These results confirm the feasibility and superiority of our CLT method for high-precision gravitational redshift testing. The approach demonstrates clear advantages over previous microwave-based methods and approaches the performance levels needed for testing alternative gravity theories as outlined by [24].

### *4.5 Result and Precision analysis*

We designed the simulation experiment to address the complex challenges of space-to-ground laser time transfer while maximizing measurement precision. To begin, we meticulously planned our observations using CSS ephemeris data, identifying optimal windows when the space station would pass over our XLS facility with elevation angles above 15°. During these passes, our ground station generated a 532 nm laser signal referenced to our optical clock and transmitted it to CSS through our telescope system. After receiving this uplink signal, CSS compared it with its onboard optical clock and reflected a return signal back to Earth. Our ground station then captured this downlink signal using single-photon detectors and time interval counters synchronized to our ground optical clock, enabling time-of-flight measurements with remarkable sub-picosecond precision.

Our data processing strategy extends significantly beyond the methods employed in previous studies [14,17]. Each observation arc typically yielded about 300 seconds of continuous measurements. We first applied quality control filters to eliminate outliers resulting from signal interruptions or detection anomalies. We then systematically corrected for known effects: geometric distance, relativistic effects, tropospheric delays (using GPT3/VMF3 models), solid Earth tides, and hardware delays. Using equation (1), we constructed our fundamental observables from these corrected measurements and applied equation (26) to eliminate remaining error terms. Instead of Sun's simple averaging approach, we developed a weighted averaging method based on the inverse square of each observation arc's standard deviation, excluding any observations that deviated by more than $3\sigma$.

Figure 3 demonstrates the exceptional performance of the simulated optical clocks. The Allan deviation plot shows that at a 1-second averaging time, our CSS space station clock (orange line) achieved stability around $2 \times 10^{-15}$, while our Xi'an ground clock (blue line) reached about $1 \times 10^{-15}$. With longer integration times, both clocks showed impressive improvement—reaching approximately $1 \times 10^{-17}$ and $7 \times 10^{-18}$ at $10^4$ seconds, and ultimately both achieving extraordinary stability around $4 \times 10^{-18}$ at nearly $10^6$ seconds. This exceptional frequency stability formed the backbone of our gravitational redshift test, contributing only about $1 \times 10^{-8}$ uncertainty to the $\alpha$ parameter.

Several factors affected the precision of our CLT-based gravitational redshift testing. Propagation medium effects proved to be our most significant challenge, with tropospheric delay variations contributing roughly $2 \times 10^{-7}$ uncertainty and atmospheric turbulence adding path length variations of about $1 \times 10^{-7}$. We also faced relativistic modeling uncertainties from orbit determination limitations and gravitational model fidelity issues, which added approximately $5 \times 10^{-8}$ and $3 \times 10^{-8}$ to our uncertainty budget, respectively.

Figure 5 showcases our precision $\alpha$ and offset measurements across 142 observation arcs. The weighted mean demonstrates remarkable consistency, with an average offset of about $1.8 \times 10^{-5}$. The error bars reveal variations that primarily reflect changing atmospheric conditions and observation geometry. The narrow confidence interval bands highlight the statistical precision we achieved through our multi-arc strategy. Ultimately, our testing methodology reached a precision of $(1.8 \pm 47) \times 10^{-7}$—a substantial improvement over previous studies. Our Allan deviation analysis of the residuals revealed white frequency noise at short integration times and flicker frequency noise at longer integration times, confirming the effectiveness of our systematic error correction approach.

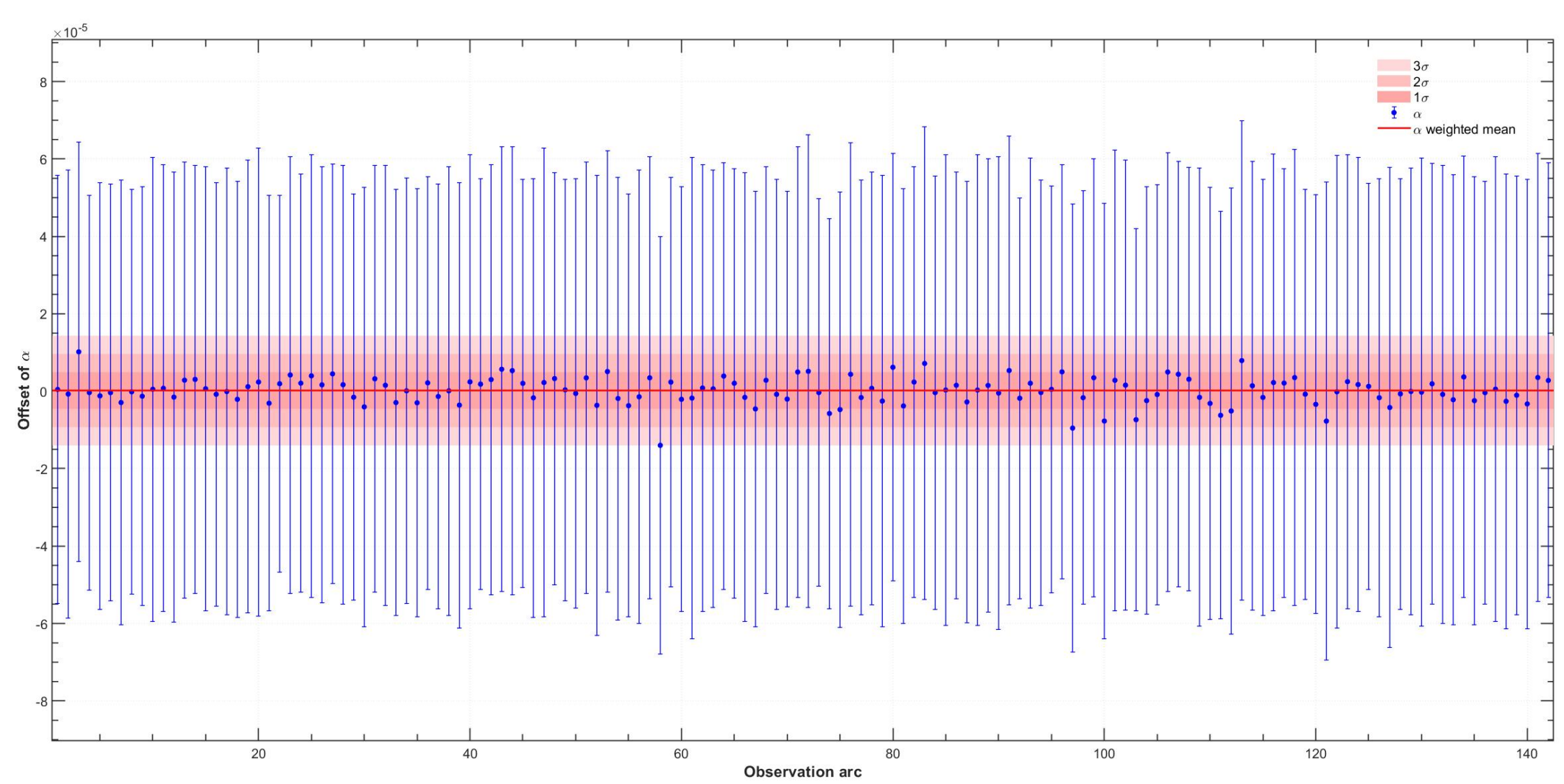

**Figure 5. Precision α and offset of CLT testing for gravitational redshift**

Our results represent a significant leap forward compared to previous gravitational redshift testing methods. For context, [14] reported a precision of $(1.0 \pm 4.9) \times 10^{-6}$ using their tri-frequency combination method with microwave links. The stability of our results throughout our 30-day observation campaign validates the reproducibility of our approach. Beyond the implications for fundamental physics, our precision enables valuable geodetic applications, including gravitational potential difference measurements with 0.1 $m^2/s^2$ precision. Looking ahead, we believe we can further enhance precision by implementing more advanced tropospheric delay models and extending our observation campaign, potentially pushing the final uncertainty down to the $10^{-8}$ level. The comprehensive error assessment and mitigation strategies implemented in our experiment establish a new benchmark for precision in gravitational redshift testing and demonstrate the potential of space-based optical time transfer for both fundamental physics investigations and practical geodetic applications.

## 5 Conclusion

Our research demonstrates the exceptional potential of the CLT system for high-precision gravitational redshift testing. Through careful simulation and analysis, we've achieved a verification precision of $(1.8 \pm 47) \times 10^{-7}$ for Einstein's gravitational redshift prediction—marking a significant advancement over previous methods. This precision stems from our comprehensive $c^{-3}$ order relativistic model and sophisticated error mitigation strategies tailored specifically for optical time transfer.

The transition to laser-based technology offers fundamental advantages that we have effectively harnessed in this study. Unlike microwave-based approaches, our 532 nm laser signals provide inherently higher frequency precision and experience negligible ionospheric effects—eliminating a major error source that plagued earlier studies. The naturally collimated laser beams significantly reduce multipath interference, while our two-way measurement configuration creates nearly symmetrical paths that enable more effective common-mode error cancellation.

Our detailed error analysis revealed that tropospheric delay and atmospheric turbulence remain the primary challenges, despite our implementation of advanced GPT3 grid modeling and VMF3 mapping functions. These atmospheric effects contribute uncertainties of approximately $2 \times 10^{-7}$ and $1 \times 10^{-7}$ respectively. Nevertheless, our systematic error correction approach has proven remarkably effective—improving link stability by nearly six orders of magnitude at longer integration times.

When comparing our results with previous work, the advantages become clear. [14] reported a precision of $(1.0 \pm 4.9) \times 10^{-6}$ using their tri-frequency microwave approach, which required three separate frequency links and still struggled with residual stabilities no better than $10^{-14}$. Our single-wavelength laser method achieves residual stabilities below $10^{-17}$ for most effects after just 1000 seconds of averaging—a substantial improvement that directly translates to more precise gravitational redshift testing.

The consistency of our results throughout our 30-day simulation campaign validates the robustness of our approach. Beyond fundamental physics, the precision we've achieved enables valuable geodetic applications, including gravitational potential difference measurements with 0.1 $m^2/s^2$ precision—facilitating intercontinental height transfer with unprecedented efficiency.

Looking forward, we see clear pathways to further enhance precision. Implementing more sophisticated tropospheric delay models that incorporate real-time meteorological data could significantly reduce our largest error source. Extended observation campaigns would allow better averaging of atmospheric effects, potentially pushing our final uncertainty down to the $10^{-8}$ level. As the CSS optical clock system completes its orbital debugging phase, we anticipate transitioning from simulation to actual experimental verification, further refining our techniques and potentially uncovering subtle deviations from Einstein's theory.

This work establishes a new benchmark for precision in gravitational redshift testing and demonstrates the transformative potential of space-based optical time transfer. The methodology we've developed provides a robust framework for future high-precision tests of General Relativity and opens exciting new possibilities for both fundamental physics investigations and practical geodetic applications.

**Acknowledgements**

This work is supported by the National Natural Science Foundation of China [Grant numbers 42388102, 42030105, 42274011, 42404005), Space Station Project [Grant numbers 2020-228], Natural Science Foundation of Henan Province [Grant Number 252300420875], Postdoctor Project of Hubei Province [Grant Number 2024HBBHCXB060], China Postdoctoral Science Foundation [Grant Number 2024M752480].

**Data availability statement**

This manuscript has no associated data or the data will not be deposited. [Authors' comment: The simulation datasets used in this study are available on reasonable request from the corresponding author(wbshen@sgg.whu.edu.cn).]

---

## Code availability statement

This manuscript has no associated code/software. [Authors' comment: Code/Software sharing not applicable to this article as no code/software was generated or analysed during the current study.]

## Competing Interests

The authors have no relevant financial or non-financial interests to disclose.

## Open Access

Funded by SCOAP$^3$.

## References

1. A. Einstein, Sitzungsberichte Der Königlich Preußischen Akademie Der Wissenschaften 844 (1915).
2. J. C. Hafele and R. E. Keating, Science **177**, 168 (1972).
3. R. F. C. Vessot and M. W. Levine, Gen Relat Gravit **10**, 181 (1979).
4. R. F. C. Vessot, M. W. Levine, E. M. Mattison, E. L. Blomberg, T. E. Hoffman, G. U. Nystrom, B. F. Farrel, R. Decher, P. B. Eby, C. R. Baugher, J. W. Watts, D. L. Teuber, and F. D. Wills, Phys. Rev. Lett. **45**, 2081 (1980).
5. P. Delva, N. Puchades, E. Schönemann, F. Dilssner, C. Courde, S. Bertone, F.

Gonzalez, A. Hees, Ch. Le Poncin-Lafitte, F. Meynadier, R. Prieto-Cerdeira, B. Sohet, J. Ventura-Traveset, and P. Wolf, Phys. Rev. Lett. **121**, 231101 (2018).

6. S. Herrmann, F. Finke, M. Lülf, O. Kichakova, D. Puetzfeld, D. Knickmann, M. List, B. Rievers, G. Giorgi, C. Günther, H. Dittus, R. Prieto-Cerdeira, F. Dilssner, F. Gonzalez, E. Schönemann, J. Ventura-Traveset, and C. Lämmerzahl, Phys. Rev. Lett. **121**, 231102 (2018).

7. M. Takamoto, I. Ushijima, N. Ohmae, T. Yahagi, K. Kokado, H. Shinkai, and H. Katori, Nat. Photonics **14**, 411 (2020).

8. C. O. Alley, in *Quantum Optics, Experimental Gravity, and Measurement Theory*, edited by P. Meystre and M. O. Scully (Springer US, Boston, MA, 1983), pp. 363–427.

9. P. Fridelance and C. Veillet, Metrologia **32**, 27 (1995).

10. P. Guillemot, K. Gasc, I. Petitbon, E. Samain, P. Vrancken, J. Weick, D. Albanese, F. Para, and J. -m. Torre, in *2006 IEEE International Frequency Control Symposium and Exposition* (Miami, 2006), pp. 771–778.

11. U. Schreiber, I. Prochazka, P. Lauber, U. Hugentobler, W. Schafer, L. Cacciapuoti, and R. Nasca, in *2009 IEEE International Frequency Control Symposium Joint with the 22nd European Frequency and Time Forum* (IEEE, Besancon, France, 2009), pp. 594–599.

12. W. Meng, H. Zhang, P. Huang, W. Jie, and I. Prochazka, Advances in Space Research **51**, 951 (2013).

13. R. Geng, Z. Wu, Y. Huang, Z. Cheng, R. Yu, K. Tang, H. Zhang, W. Meng, H. Deng, M. Long, S. Qin, and Z. Zhang, Advances in Space Research **73**, 2548 (2024).

14. X. Sun, W.-B. Shen, Z. Shen, C. Cai, W. Xu, and P. Zhang, Eur. Phys. J. C **81**, 634 (2021).

15. Z. Shen, W. Shen, T. Zhang, L. He, Z. Cai, X. Tian, and P. Zhang, Advances in Space Research **68**, 2776 (2021).

16. China Manned Space Engineering Office, 38 (2019).

17. W. Shen, P. Zhang, Z. Shen, R. Xu, X. Sun, M. Ashry, A. Ruby, W. Xu, K. Wu, Y. Wu, A. Ning, L. Wang, L. Li, and C. Cai, Phys. Rev. D **108**, 064031 (2023).

18. L. Blanchet, C. Salomon, P. Teyssandier, and P. Wolf, A&A **370**, 320 (2001).

19. J. Müller, M. Soffel, and S. A. Klioner, J Geod **82**, 133 (2008).

20. M. Soffel, *Space-Time Reference Systems* (Springer, 2013).

21. P. Exertier, E. Samain, N. Martin, C. Courde, M. Laas-Bourez, C. Foussard, and P. Guillemot, Advances in Space Research **54**, 2371 (2014).

22. E. Samain, P. Exertier, C. Courde, P. Fridelance, P. Guillemot, M. Laas-Bourez, and J. M. Torre, Metrologia **52**, 423 (2015).

23. D. Kleppner, R. F. C. Vessot, and N. F. Ramsey, Astrophys Space Sci **6**, 13 (1970).

24. C. M. Will, Living Rev. Relativ. **17**, 4 (2014).

25. E. Savalle, C. Guerlin, P. Delva, F. Meynadier, C. le Poncin-Lafitte, and P. Wolf, Classical and Quantum Gravity **36**, 245004 (2019).

26. D. Dirkx, R. Noomen, P. N. A. M. Visser, S. Bauer, and L. L. A. Vermeersen, Planetary and Space Science **117**, 159 (2015).

27. C. Hohenkerk, in *The Science of Time 2016*, edited by E. F. Arias, L. Combrinck, P. Gabor, C. Hohenkerk, and P. K. Seidelmann (Springer International Publishing, Cham, 2017), pp. 159–163.

28. G. Petit and B. Luzum, IERS Technical Note **36**, 1 (2010).

29. I. I. Shapiro, Phys. Rev. Lett. **13**, 789 (1964).

30. J. W. Marini, Radio Science **7**, 223 (1972).

31. V. B. Mendes, G. Prates, E. C. Pavlis, D. E. Pavlis, and R. B. Langley, Geophysical Research Letters **29**, 53 (2002).

32. T. Nilsson, J. Böhm, D. D. Wijaya, A. Tresch, V. Nafisi, and H. Schuh, *Path Delays in the Neutral Atmosphere* (2013), p. 136.

33. D. Landskron and J. Böhm, J Geod **92**, 349 (2018).

34. L. Duchayne, F. Mercier, and P. Wolf, A&A **504**, 653 (2009).

35. Meng W., Research on Laser Time Transfer and Payload Technology for Chinese Space Station, Doctoral dissertation, East China Normal University, 2021.

36. J. Degnan, in *Proceedings 20th ILRS Workshop, Potsdam, Germany* (2016).

37. L. C. Andrews and R. L. Phillips, *Laser Beam Propagation through Random Media* (SPIE, 1000 20th Street, Bellingham, WA 98227-0010 USA, 2005).

38. G. I. Taylor, Proc. R. Soc. Lond. A **164**, 476 (1938).

39. M. T. Taylor, A. Belmonte, L. Hollberg, and J. M. Kahn, Phys. Rev. A **101**, 033843 (2020).

40. P. Fridelance, E. Samain, and C. Veillet, Experimental Astronomy **7**, 191 (1997).

41. J. Wahr, in *Global Earth Physics* (1995), pp. 40–46.

42. C. Voigt, H. Denker, and L. Timmen, Metrologia **53**, 1365 (2016).

43. J. A. Barnes, A. R. Chi, L. S. Cutler, D. J. Healey, D. B. Leeson, T. E. McGunigal, J. A. Mullen, W. L. Smith, R. L. Sydnor, R. F. C. Vessot, and G. M. R. Winkler, IEEE Transactions on Instrumentation and Measurement **IM–20**, 105 (1971).

44. C. Zucca and P. Tavella, IEEE Transactions on Ultrasonics, Ferroelectrics, and Frequency Control **52**, 289 (2005).

45. C. Zucca and P. Tavella, Metrologia **52**, 514 (2015).

46. N. J. Kasdin, Proceedings of the IEEE **83**, 802 (1995).

47. P. Lesage and C. Audoin, IEEE Transactions on Instrumentation and Measurement **22**, 157 (1973).

48. S. R. Stein, in *Precision Frequency Control* (Academic Press, New York, 1985), pp. 191–416.

49. P. Lesage and C. Audoin, Radio Science **14**, 521 (1979).

50. N. Pavlis, S. Holmes, S. Kenyon, and J. Factor, in (2008), pp. G22A-01.

51. P. Wolf and G. Petit, Astronomy and Astrophysics **304**, 653 (1995).

52. K. U. Schreiber, T. Klügel, J.-P. R. Wells, R. B. Hurst, and A. Gebauer, Phys. Rev. Lett. **107**, 173904 (2011).

53. P. Exertier, A. Belli, and J. M. Lemoine, Advances in Space Research **60**,

948 (2017).

54. M. Soffel, S. A. Klioner, G. Petit, P. Wolf, S. M. Kopeikin, P. Bretagnon, V. A. Brumberg, N. Capitaine, T. Damour, T. Fukushima, B. Guinot, T.-Y. Huang, L. Lindegren, C. Ma, K. Nordtvedt, J. C. Ries, P. K. Seidelmann, D. Vokrouhlický, C. M. Will, and C. Xu, The Astronomical Journal **126**, 2687 (2003).

55. N. Ashby and B. Bertotti, Phys. Rev. D **34**, 2246 (1986).

56. F. Riehle, Nature Photonics **11**, 25 (2017).

57. P. Delva, J. Lodewyck, S. Bilicki, E. Bookjans, G. Vallet, R. Le Targat, P.-E. Pottie, C. Guerlin, F. Meynadier, C. Le Poncin-Lafitte, O. Lopez, A. Amy-Klein, W.-K. Lee, N. Quintin, C. Lisdat, A. Al-Masoudi, S. Dörscher, C. Grebing, G. Grosche, A. Kuhl, S. Raupach, U. Sterr, I. R. Hill, R. Hobson, W. Bowden, J. Kronjäger, G. Marra, A. Rolland, F. N. Baynes, H. S. Margolis, and P. Gill, Phys. Rev. Lett. **118**, 221102 (2017).

58. G. Petit and P. Wolf, Metrologia **42**, S138 (2005).

59. C. Lisdat, G. Grosche, N. Quintin, C. Shi, S. M. F. Raupach, C. Grebing, D. Nicolodi, F. Stefani, A. Al-Masoudi, S. Dörscher, S. Häfner, J.-L. Robyr, N. Chiodo, S. Bilicki, E. Bookjans, A. Koczwara, S. Koke, A. Kuhl, F. Wiotte, F. Meynadier, E. Camisard, M. Abgrall, M. Lours, T. Legero, H. Schnatz, U. Sterr, H. Denker, C. Chardonnet, Y. Le Coq, G. Santarelli, A. Amy-Klein, R. Le Targat, J. Lodewyck, O. Lopez, and P.-E. Pottie, Nature Communications **7**, 12443 (2016).